\documentstyle[preprint,aps]{revtex}

\begin{document}
\draft
\preprint{}
\title{Degeneracy Algorithm for Random Magnets}
\author{S. Bastea\cite{sb}}
\address{Department of Physics $\&$ Astronomy and Center for Fundamental 
Materials Research,\\Michigan State University,
East Lansing, Michigan 48824-1116}
\maketitle
\begin{abstract}
It has been known for a long time that the ground state problem of 
random magnets, e.g. random field Ising model (RFIM), can be mapped 
onto the max-flow/min-cut problem of 
transportation networks. I build on this approach, relying on the 
concept of residual graph, and design an algorithm that I prove to 
be exact for finding all the minimum cuts, i.e. the ground state 
degeneracy of these systems. I demonstrate that this algorithm is 
also relevant for the study of the ground state properties of the dilute Ising 
antiferromagnet in a constant field (DAFF) and interfaces in random 
bond magnets.
\end{abstract}

\widetext
\section{Introduction}
The statistical physics of random and frustrated systems has received a
considerable deal of attention in recent years \cite{young}. 
The presence of quenched disorder has been found to greatly change the 
bulk and interface properties of a variety of systems as compared with 
their 'pure' counterparts, leading to new and very interesting equilibrium 
and nonequilibrium phenomena. Unfortunately the progress has been many times 
rather slow, primarily because random systems pose sometimes insurmountable 
theoretical difficulties even to the most stubborn theorists.
Computer simulations have played and continue to play an important 
role in the field, 
being at times the only guide through a very complicated energy landscape.
While the traditional Monte Carlo method \cite{binder}
proved its usefulness again and again, it was soon realized that other 
approaches should be considered, depending on the nature of the problem at 
hand. Since then a variety of algorithms, previously known only within 
the computer science community, have been successfully brought to bear on 
numerous statistical mechanics problems with quenched disorder, from spin 
glasses \cite{barahona} to rigidity percolation \cite{duxbury1}. 
Such algorithms, generally known as 
combinatorial optimization algorithms, have been typically used 
to find the exact $T=0$ ground states of the system being studied, 
completely avoiding the equilibration problems specific to the Monte 
Carlo simulations.

In the following I will focus on a single class of such algorithms, 
network flow algorithms, that have been put in the limelight by 
the work of Ogielski \cite{ogielski}, who applied them to the study 
of the random field Ising model (RFIM). Since then the same method was 
also successfully applied to the study of equilibrium interfaces in 
disordered systems \cite{middleton,duxbury2}, becoming 
an important tool for the physicists working in the field.
The method is generally based upon mapping the system being studied 
onto a network of capacities through which an incompressible  
fluid obeying local 
mass conservation flows. The problem of finding the ground state turns 
out to be equivalent to finding the maximum flow that can be pushed 
through the network between two special nodes, the source and the sink, 
the so called max-flow/min-cut problem of operations research
\cite{ford,comb,vlsi}. 
The advantage of this approach is that polynomial time algorithms 
have been developed for this problem\cite{comb,vlsi,goldberg}, some 
of which are much older than the field of random systems.

One of the problems that was usually not addressed using these algorithms 
and an important problem in the physics of random systems is the 
degeneracy of the $T=0$ ground states. In terms of the associated network 
flow problem this is the question of minimum cut degeneracy.
An approximate algorithm dealing with this issue was proposed 
in \cite{hartmann} and applied to the RFIM problem.
In this paper I build on the max-flow/min-cut approach,
relying strongly on the concept of residual graph \cite{goldberg},
and design an exact algorithm for finding all the minimum-cuts 
(or equivalently all the ground states for a certain class of systems).

The organization of the paper is the following: for the sake of completeness 
Section II introduces the mapping of the ground state problem to 
the max-flow/min-cut problem along with the network flow terminology and 
two combinatorial results that will be used in the design of the algorithm, 
Section III describes the degeneracy algorithm, Section IV presents a number  
of applications of the algorithm and the Appendix includes the proofs to 
Propositions 1-4. 

\section{Ground states using max-flow/min-cut algorithms}
In the following I will present the mapping of the ground state 
problem to the max-flow/min-cut problem and then proceed to describe what 
kind of information one can extract from this mapping. I will use the  
$T=0$ interface problem in the random bond Ising ferromagnet to illustrate the 
method because it is somewhat easier for the unfamiliar reader to 
understand it intuitively ( see also \cite{rieger} for a review). 

The Hamiltonian of the system is
\begin{equation}
H=-\sum_{\langle ij \rangle} J_{ij}\sigma_i\sigma_j
\end{equation}
where $\sigma_i = \pm 1$ and $J_{ij}\geq 0$ are ferromagnetic couplings between
neighboring spins. $J_{ij}$ are fixed independent identically distributed 
random variables - quenched randomness. If the system is a $d$-dimensional 
cube, an interface with dimension $d-1$ can be induced by using periodic 
boundary conditions along $d-1$ directions and setting the spins in the 
two $d-1$ dimensional hyperplanes that represent the boundaries of the 
system along the last direction to $+1$ and $-1$, respectively. The 
interface that will form in the system between the $+1$ and $-1$ hyperplanes
 will generally be rough, 
because it will wander in order to break the weakest bonds. The energy is a 
minimum over all the possible spin configurations (with the $\pm1$ 
boundary spins fixed) and therefore the problem of finding the minimum energy 
configuration(s) would appear to be computationally very hard,
even for small system sizes. As it turns out, this is not the case if one
takes advantage of the similarity between this problem and the 
max-flow/min-cut problem, very well known in the study of transportation 
networks \cite{ford,comb,vlsi}. The idea is the following:
two new extra sites are introduced, a source node $s$ that is connected to all
the spins of the $+1$ hyperplane, and a sink node $t$ connected to all
the spins of the $-1$ hyperplane. The ferromagnetic constants coupling the 
source node $\sigma_s$ and the sink node $\sigma_t$ to their corresponding 
hyperplane are chosen to be strong enough so they are not broken in the 
ground state. Then 
by setting $\sigma_s = +1$ and $\sigma_t=-1$ an interface is induced as before.
Now we view the system, including $s$ and $t$,  as a graph whose arcs are the 
bonds between the spins. The arcs have forward and backward capacities equal 
with the corresponding coupling constants, $c_{ij} = c_{ji} = J_{ij}$, 
or we can imagine that the nodes $i$ and $j$ are connected by both forward 
and backward arcs with capacities $c_{ij}$ and $c_{ji}$, so this 
is a directed graph.( The constraint that $c_{ij} = c_{ji}$  
is dictated by the physics and is not specific to the general 
max-flow/min-cut problem. It can be relaxed for the interface 
problem, but not for the random field problem.) We define the set of nodes 
as $N$ and a partition $(X, Y)$ of the nodes as
\begin{equation}
X\equiv \{i\in N|\sigma_i = +1\}
\end{equation}
\begin{equation}
Y\equiv \{i\in N|\sigma_i = -1\}
\end{equation}
Then $X\cup Y = N$, $X\cap Y = \emptyset$, $s\in X$ and $t\in Y$.
 The knowledge of such a partition determines the energy of the spin 
configuration and the position of the interface. This is readily seen if 
we write the Hamiltonian of the system as
\begin{equation}
H=-\sum_{(i,j)\in A(X)} J_{ij}-\sum_{(i,j)\in A(Y)} 
J_{ij}+\sum_{(i,j)\in A(X, Y)} J_{ij}=
H_0 + 2\sum_{(i,j)\in A(X, Y)}J_{ij}
\end{equation}
where $H_0$ is the energy of the fully aligned system, 
$H_0 = \sum_{(i,j)}J_{ij}$, 
and we defined $A(X) \equiv\{(i,j)|i\in X, j\in X\}$, 
$A(Y) \equiv\{(i,j)|i\in Y, j\in Y\}$ and  
$A(X, Y) \equiv\{(i,j)|i\in X, j\in Y\}$. Thus the problem of finding the 
ground state interface, which has minimum energy, is equivalent to 
finding a partition $(X, Y)$, also called a cut, that minimizes 
$\sum_{(i,j)\in A(X, Y)}J_{ij}$ - {\it minimum cut}.
(Note that $H_0$ is a constant for a given random sample.)
If we imagine fluid flowing through the network from the source $s$ to the 
sink $t$, the minimum cuts are the bottlenecks of the network of 
capacities because they determine the maximum flow that can be 
pushed through the network from $s$ to $t$.
It is useful at this point to put the problem in a more general setting. 
For this a number of definitions are in order.

A {\it transportation network} is a directed graph $G_c(N,A)$ with two 
special nodes, $s$ - source and $t$ - sink or target; 
$N$ is the set of nodes and $A$ the set of arcs. The directed arc $(ij)$ 
going from node $i$ to $j$ has capacity $c_{ij}\geq0$. (For the sake of 
clarity we assume that if the arc $(ij)$ exists and has capacity $c_{ij}>0$ 
then $(ji)$ also exists and has capacity $c_{ji}\geq 0$.)
A {\it flow} through the network $G_c(N,A)$ is a set of numbers $\{f_{ij}\}$, 
each corresponding to an arc in $A$, subject to the following feasibility 
constraints:
\begin{eqnarray}
&&0\leq f_{ij}\leq c_{ij}\hspace{0.75cm}\mbox{capacity constraint}\\
&&e_i=\sum_{\{j|(ji)\in A\}}f_{ji}-\sum_{\{j|(ij)\in A\}}f_{ij}=0
\hspace{0.75cm}\mbox{local flow conservation}
\end{eqnarray}
for all the nodes $j\in N-\{s,t\}$, and 
\begin{eqnarray}
-e_s=e_t=f
\end{eqnarray}
where $f$ is called the value of the flow. (Note that $-e_s=e_t$ follows 
from the flow conservation \cite{comb}.)
The {\it maximum flow problem} of network flows
is concerned with finding the flow $\{f_{ij}\}$ through the 
transportation network $G_c$ that has maximum value $f$.

For any feasible flow through the network we define the {\it residual graph} 
as the graph $G_f(N,A)$ with positive {\it residual capacities} of the arcs,
\begin{eqnarray}
r_{ij}=c_{ij}-f_{ij} + f_{ji}>0
\end{eqnarray}
where $f_{ij}-f_{ji}$ is the net flow from $i$ to $j$. (Note that 
$r_{ij}\geq 0$ follows from the capacity constraint, and also 
that it is possible that $c_{ij}=0$ and $r_{ij}>0$, when $c_{ji}>0$.)
An {\it augmenting path} is a directed path from $s$ to $t$ in the 
residual graph $G_f$.

A {\it cut} is a partition of the nodes set $N$ 
into two subsets $X$ and $Y$, denoted by $(X, Y)$, with $s\in X$ and 
$t\in Y$, such that $X\cup Y = N$ 
and $X\cap Y = \emptyset$.
The {\it capacity of a cut} is defined as: 
\begin{eqnarray}
c(X,Y)=\sum_{\{(ij)\in A|i\in X, j\in Y\}}c_{ij}
\end{eqnarray}

In the following I will concentrate on the case of a transportation network 
$G_c(N,A)$ in which if the arc $(ij)$ exists and has capacity $c_{ij}>0$ 
than the arc $(ji)$ also exists 
and has capacity $c_{ji}>0$. This kind of network is the most relevant one 
in the physics applications that I described before. For these networks the 
following two propositions can be proven (the proofs are contained in the 
Appendix), that can be used to design an 
algorithm that finds all the minimum cuts in the network $G_c(N,A)$.
 
\underline{{\it Proposition 1}}: If $\{f_{ij}\}$ is a maximal flow and 
$(X,Y)$ a minimum cut then $r_{ij}=0$ for all arcs 
$\{(ij)\in A|i\in X, j\in Y\}$.

\underline{{\it Proposition 2}}: If $f_{max}>0$ a cut $(X,Y)$ in $G_c$ 
is a minimum cut if and only if it is a directed partition in $G^{max}_f$, 
the residual graph for a maximal flow. 

The calculation of an actual maximal flow
through the network can be done using polynomial time algorithms.
The first such algorithm, the {\it augmenting path algorithm} \cite{comb},
 was proposed 
originally by Ford and Fulkerson and it is also a way to prove the 
max-flow/min-cut theorem (see Appendix). However, much faster algorithms 
have been developed 
in recent years, in particular push-relabel methods with global 
updates, that allow one to deal with much bigger systems than before 
\cite{vlsi}.
These algorithms can be and have been used in such a way as 
to improve the computational speed by providing only the value of 
the maximal flow and a minimum cut, but not an actual maximal 
flow through the network \cite{goldberg}. However, the knowledge of an 
actual maximal flow through the network is crucial for finding all 
the minimum cuts. Fortunately, such a calculation can be made without a 
major loss of speed \cite{goldberg}.

\section{A degeneracy algorithm}
In the following I will use Propositions 1 and 2 to 
design an algorithm that finds all the minimum cuts. The algorithm 
will be aimed at finding the set of arcs $\{(ij)\in A|i\in X, j\in Y, 
\mbox{{\it for all minimum cuts }(X, Y)}\}$, denoted thereafter by 
${\cal A}_{mc}$, and a procedure for determining the actual minimum cuts.

Let us assume that we constructed a maximal flow $\{f^{max}_{ij}\}$ 
through the network $G_c(N, A)$ using an appropriate algorithm and 
let $G_f^{max}$ be the associated residual graph.
We define:
\begin{equation}
Y_t\equiv\{\mbox{{\it all nodes that can reach the sink along a directed 
path in }}G_f^{max}\}
\end{equation}
\begin{equation} 
X_t=N-Y_t
\end{equation}
Then $(X_t,Y_t)$ is minimum cut that we will call
the $T$ cut (this follows from Proposition 2). This minimum cut has 
the property that for any other minimum cut $(X,Y)$, $Y\cap Y_t=Y_t$. 
({\it Proof}: Let us assume $Y\cap Y_t=Y_1\neq Y_t$; 
then $X_1=Y_t-Y_1\subset X$. Because  
$(X,Y)$ is a directed cut (Proposition 2), for any node $i\in X$ there 
is no directed path 
from $i$ to the sink $t$ in $G^{max}_f$, which must also be true for any 
$i\in X_1$. But $X_1\subset Y_t$, so this is a contradiction.)
We also define:
\begin{equation}
X_s\equiv\{\mbox{{\it all nodes that can be reached from 
the source along a directed path in }} G^{max}_f\} 
\end{equation}
\begin{equation}
Y_s=N-X_s
\end{equation}
Then $(X_s,Y_s)$ is also a minimum cut 
that we call the $S$ cut. This minimum cut has the property that 
for any other minimum cut $(X,Y)$, $X\cap X_s=X_s$.
Now we define
\begin{equation}
Z=N-X_s-Y_t
\end{equation}
Then if $Z=\emptyset$ a single minimum cut 
exists, otherwise there are at least two. We will be concerned with the 
non-trivial case $Z\neq\emptyset$.

At this point our knowledge of ${\cal A}_{mc}$ is summarized in 
Fig. \ref{1step}. The arrows stand for possibly more than one arc of 
$G(N,A)$, and all these arcs are included in ${\cal A}_{mc}$. 
(Note that these arcs are saturated by the flow, that is $r_{ij}=0$, 
so in $G^{max}_f$ only the arcs $(ji)$ are present). By the construction 
of $Z$ we also know that all the remaining arcs that make up ${\cal A}_{mc}$
connect nodes that are included in $Z$.

The first part of the algorithm is aimed at finding the disconnected pieces 
(clusters) that make up $Z$, and is therefore called {\it cluster counting}.
A look at Fig. \ref{rfim}, which is an application of the algorithm to 
the bimodal RFIM, clarifies the significance of this step.
The idea is to determine the connectivity of $Z$ in $G_f^{max}$
using as connectivity rule $(r_{ij}\neq 0\mbox{ or } r_{ji}\neq 0)$ 
, which is equivalent to $(c_{ij}\neq 0\mbox{ or } c_{ji}\neq 0)$. 
This procedure does not reveal any new arcs of ${\cal A}_{mc}$, 
but shows how the minimum cuts are constructed 
using the arcs of ${\cal A}_{mc}$ that we know at this moment. 
The well known Hoshen-Kopelman algorithm \cite{stauffer}, which is a an  
efficient cluster labeling procedure, can be readily 
adapted for the cluster counting task, and allows such a calculation to 
be made in a time proportional with the number of nodes in $Z$ \cite{amend}. 
Fig. \ref{2step} summarizes our knowledge of the minimum cuts 
after this step. It is easy to see that the number of minimum 
cuts that we can construct at this time is $2^{n(Z)}$, where $n(Z)$ is 
the number of independent clusters making up $Z$. Our search 
for the remaining minimum cuts is then reduced to finding the minimum 
cuts in each of the independent clusters. If $f_{max}=0$ the
algorithm can be stopped here as everything is known about the minimum cuts. 
Any partition of the $n(Z)$ clusters is a minimum cut and the total number 
of minimum cuts is $2^{n(Z)}$. (Note that if $f_{max}>0$ the 'zero' step 
of the algorithm should be to find the cluster percolating from $s$ to $t$ 
with the above connectivity rule. All the other clusters contribute 
in a trivial manner to the minimum cuts, therefore in the 
following we assume that they have been already discarded.)

In the second part of the algorithm we perform a {\it subcluster counting}
procedure on each of the independent clusters found at the first step. 
The connectivity rule that we use is $(r_{ij}\neq 0 \mbox{ and } 
r_{ji}\neq 0)$. It is clear, using Proposition 1, that the subclusters
so obtained can only be on one side or the other of a minimum cut, 
i.e. the set ${\cal A}_{mc}$ is included in the set of arcs  
that connect these subclusters to one another (we formally call $X_s$ and 
$Y_t$ subclusters). In order to eliminate 
the overcounting, we apply a third procedure called 
{\it subcluster coagulation}. 

At the end of the second step of the algorithm the subclusters are connected 
with each other possibly through multiple arcs with $r_{ij}=0$ and $r_{ji}>0$. 
The {\it subcluster coagulation} procedure is applied iteratively on each 
of the independent clusters and consists of the coagulation of 
subclusters that make up a directed cycle. A directed cycle is an ordered 
sequence of subclusters, $S_1$, $S_2$, ..., $S_m$, such that for all $S_k$, 
$k=1, ..., m$, exist $i_k$, $j_k$ nodes, $i_k,j_k\in S_k$, not necessarily 
distinct, with the property that $r_{i_kj_{k+1}}=0$ and $r_{j_{k+1}i_k}>0$, 
where $S_{m+1}\equiv S_1$. The idea is that, according to 
Proposition 2, subclusters making up a such directed cycle 
cannot be on different parts of a minimum cut, therefore the arcs connecting 
them are not included in ${\cal A}_{mc}$ (see also the discussion following 
Proposition 4). A Hoshen-Kopelman
type algorithm, in which the subcluster coagulation is achieved through 
relabeling of the subclusters, is again rather efficient at handling this task,
but managing the data structure requires rather intricate coding.

At the end of the algorithm we have constructed a supergraph like the one
in Fig. \ref{3step}, which we denote by ${\cal G(N, A)}$,
with single directed arcs and no cycles. 
The nodes are now the subclusters and the  arcs stand, as before, for 
possibly more than one arc of $G(N, A)$. Formally, the arc $(IJ)$ going  
from the subcluster $I$ to subcluster $J$ is defined by:
\begin{equation}
(IJ)=\{(ij)\in A|i\in I, j\in J\}
\end{equation}
and by the construction of the algorithm $r_{ij}=0$ and $r_{ji}>0$ for 
all arcs $(ij)\in (IJ)$. 
Then we define ${\cal A}_{sg}=\cup_{{\cal A}} (IJ)$,
the set of arcs of $G(N, A)$ represented by the arcs of ${\cal G(N, A)}$. 
The following proposition is then true (see Appendix for the proof):

\underline{{\it Proposition 3}}: At the end of the algorithm 
${\cal A}_{sg}={\cal A}_{mc}$.

When the algorithm terminates ${\cal A}_{mc}$ is known and 
moreover, the problem of counting the minimum cuts is reduced to finding all
the directed partitions in a directed, much smaller supergraph - 
${\cal G(N, A)}$ - 
with single arcs and no cycles. Simple enumeration is therefore 
feasible if the independent clusters are not too big. The degeneracy 
(total number of minimum cuts) $D$ can be written as:
\begin{eqnarray}
D=\prod_i d(i)
\end{eqnarray}
where $d(i)$ is the degeneracy associated with cluster $i$, $d(i)\ge 2$.

One additional result can be proven (the proof is contained in the Appendix), 
that was also probably known by Ford and Fulkerson 
\cite{ford}, that further clarifies the meaning of ${\cal A}_{mc}$.

\underline{{\it Proposition 4}}: ${\cal A}_{mc}$ is the set of arcs 
of $G(N,A)$ that will be saturated by all the maximal flows (the 'weak' 
links of the network).

The above proposition makes it easier to understand intuitively the 
significance of the subcluster coagulation step of the algorithm. It is clear 
that the flows between subclusters are all saturating and also that the 
flow conservation holds for each subcluster as a unit. Therefore, if for 
a certain maximal flow we identify a saturating flow cycle between 
subclusters, we can reduce the flow around the cycle by its smallest value 
between two subclusters making up the cycle, while keeping the overall 
flow maximal. This implies that none of the arcs making up the cycle is 
necessarily saturated when the flow is maximal, therefore, according to 
Proposition 4, they are not part of ${\cal A}_{mc}$. Furthermore, because 
of the way the subclusters have been constructed, none of the arcs connecting 
the subclusters making up the cycle is part of ${\cal A}_{mc}$, so the  
subclusters can be coagulated.

\section{Applications}

Our interest in designing a degeneracy algorithm arose in connection with 
our desire to study the ground state properties of random magnets.
We first used the above algorithm to study the ground state structure 
of the two-dimensional $\pm h$ random field Ising model (RFIM), that we 
expected to have a large degeneracy. The Hamiltonian of  
the RFIM is:
\begin{eqnarray}
H_{RFIM}=-J\sum_{\langle ij \rangle}\sigma_i\sigma_j - \sum_{i} h_i\sigma_i
\end{eqnarray}
$J>0$ and $h_i$'s are independent random variables drawn from a symmetric
distribution $P(h_i)$, with $\langle h_i\rangle=0$ and 
$\langle h_i^2\rangle^\frac{1}{2}=h$. In the case of the $\pm h$ RFIM the 
$h_i's$ are, with equal probability, $+h$ and $-h$. This problem is mapped 
onto the network flow problem by connecting the sites with positive 
fields to the source and the ones with negative fields to the sink 
through arcs with capacity $h$ (see for example \cite{rieger} for 
more details).

The structure of a $\pm h$ RFIM ground state is shown in Fig. \ref{rfim}:
two frozen ferromagnetic domains of 'up' and 'down' spins and a number 
of isolated clusters that can be flipped (colored) independently of each 
other to generate new ground states. There is also a degeneracy 
associated to flipping certain groups of subclusters inside a cluster.
Surprisingly, domains that can be flipped without changing the energy 
exist even if $h/J$ is irrational. In this case they have 
zero field energy and the same exchange 
energy in the 'up' and 'down' states,  and are located at the boundary of 
the frozen domains. The smallest such clusters have only two spins 
( see Fig. \ref{rfim}). As a result of this structure the $\pm h$ RFIM 
has a strictly 
positive entropy for a range of $h/J$, and related with it 
a new order parameter, the paramagnetic response associated with the 
orientation of the above domains. This result may be relevant to 
universality issues that are currently being debated \cite{aura,banavar}.  

A similar structure is found for the dilute Ising antiferromagnet 
in a constant field (DAFF), Fig. \ref{daff}, which is believed to be 
the experimental realization of the RFIM:
\begin{eqnarray}
H_{DAFF}=J\sum_{(ij)}\epsilon_i\epsilon_j\sigma_i\sigma_j - 
h\sum_{i}\epsilon_i\sigma_i
\end{eqnarray}
$J>0$ and $\epsilon_i$ are quenched independent random variables,
$P(\epsilon_i)=p\delta(\epsilon_i-1)+(1-p)\delta(\epsilon_i)$, $0<p\leq 1$.
Detailed results are being reported elsewhere \cite{bastea1}.

For the sake of clarity one of the clusters in Fig. \ref{rfim} and its 
subcluster graph representation as given by the algorithm are shown in 
Fig. \ref{cluster}. The connections to $X_s$ and $Y_t$ are not shown;
any directed partition is a minimum cut. Note that the arrows separating 
such a partition point from the 'up' subclusters to the 'down' subclusters.

We also applied the algorithm to the study of the ground state interfaces 
in the bond-diluted Ising model. The Hamiltonian is the one in 
Section II, with $J_{ij}$ being $J>0$ with probability $p$ and $0$ with 
probability $1-p$. The ground state structure is shown in Fig. \ref{int1},
with neighboring subclusters having different colors. These subclusters 
can be thought of as the excitations of a single interface, as in 
Fig. \ref{int1}, or multiple interfaces as in Fig. \ref{int2}. The size 
distribution of these excitations is a power law. Detailed 
results will be reported elsewhere \cite{bastea2}.

In conclusion, we designed an exact algorithm that finds all the 
arcs of a flow network that are part of a minimum cut and allows the 
effective construction of all the minimum cuts. This algorithm 
is relevant for the study of the 
ground state properties of the random field Ising model (RFIM), 
dilute Ising antiferromagnet in a field (DAFF), interface properties 
of certain Ising models with bond or site disorder and for other 
physical problems that can be mapped onto network flow problems and where the 
ground state is expected to be degenerate.

After this work was completed Bruce Hendrickson brought to my attention the 
work of Ball and Provan \cite{ball}, that contains the idea of constructing 
an acyclic graph that can be viewed as a compact representation of all 
minimum cuts (see also \cite{provan}). The authors also present a polynomial 
algorithm for counting the directed cuts in a planar acyclic graph. It appears 
that these problems are of continuing interest for the study of network 
reliability.

\section{Acknowledgments}
I am happy to acknowledge numerous fruitful discussions with Phillip M. 
Duxbury, whom I also thank for unwavering support. I thank Bruce Hendrickson 
for providing some relevant references. This work was funded by the U.S. DOE 
under contract DE-FG02-90ER45418.

\section{Appendix}
In the following we present the proofs of the Propositions 1-4 that were 
quoted in the main text. They rely in 
part on the max-flow/min-cut theorem of Ford and Fulkerson \cite{ford,comb}, 
probably the most important result in network flows:
 
\underline{{\it Theorem}} (Ford and Fulkerson, 1956): In a transportation 
network $G_c(N,A)$ the maximum value of $f$ over all flows $\{f_{ij}\}$ 
is equal to the minimum value $c(X,Y)$ over all cuts $(X,Y)$.

\underline{{\it Proposition 1}}: If $\{f_{ij}\}$ is a maximal flow and 
$(X,Y)$ a minimum cut then $r_{ij}=0$ for all arcs 
$\{(ij)\in A|i\in X, j\in Y\}$.

{\it Proof}: Let $\{f_{ij}\}$ be a maximal flow with value $f_{max}$ and 
$(X,Y)$ a minimum cut. We have $f=f_{max}=
\sum_{\{(ij)\in A|i\in X, j\in Y\}}(f_{ij} - 
f_{ji})$ by conservation of the flow \cite{comb}, and also 
$f_{max}=\sum_{\{(ij)\in A|i\in X, j\in Y\}}c_{ij}$, 
by the Ford-Fulkerson theorem.
This immediately implies $\sum_{\{(ij)\in A|i\in X, j\in Y\}}r_{ij}=0$, 
and because $r_{ij}\geq 0$ Proposition 1 follows.

\underline{{\it Proposition 2}}: If $f_{max}>0$ a cut $(X,Y)$ in $G_c$ 
is a minimum cut if and only if it is a directed partition in $G^{max}_f$, 
the residual graph for a maximal flow. 

{\it Definition}: A partition $(X^*, Y^*)$ of the nodes $N$, $X^*\cup Y^*=N$, 
$X^*\cap Y^*=\emptyset$, is directed if 
the arcs connecting $X^*$ and $Y^*$ all have the same direction, 
for example going from $Y^*$ to $X^*$, i.e. $r_{ij}=0$ for all arcs  
$\{(ij)\in A|i\in X^*, j\in Y^*\}$ and $\exists (j^*i^*)\in A$, 
$i^*\in X^*$, $j^*\in Y^*$, $r_{j^*i^*}>0$.

{\it Proof}: If $(X,Y)$ is a minimum cut we have from Proposition 1
$r_{ij}=0$ for all arcs $\{(ij)\in A|i\in X, j\in Y\}$. Now if we also 
assume $r_{ji}=0$ for all arcs $\{(ji)\in A|i\in X, j\in Y\}$ this implies 
$c_{ij}=c_{ji}=0$ for all arcs $\{(ij)\in A, (ji)\in A|i\in X, j\in Y\}$, 
and further that 
$f_{max}=0$ from the Ford-Fulkerson theorem, which is a contradiction. Thus 
the direct implication is proved.\\
Conversely, let us now assume that $(X^*,Y^*)$ is a directed partition in 
$G^{max}_f$, i.e. $r_{ij}=0$ for all arcs  
$\{(ij)\in A|i\in X^*, j\in Y^*\}$ and $\exists (j^*i^*)\in A$, 
$i^*\in X^*$, $j^*\in Y^*$, $r_{j^*i^*}>0$. In order to prove that 
$(X^*,Y^*)$ is a minimum cut we have to prove first that $(X^*,Y^*)$ is 
a cut, i.e. $s\in X^*$ and $t\in Y^*$. Let us assume that this is not 
true. Then either $a)(s\in Y^*$, $t\in X^*)$, or $b)(s\in X^*$, $t\in X^*)$, 
or $c)(s\in Y^*$, $t\in Y^*)$. We will show that all these lead 
to a contradiction.\\
$a)(s\in Y^*$, $t\in X^*)$. In this case $(Y^*,X^*)$ is a cut and 
by the conservation of the flow 
$\sum_{\{(ij)\in A|i\in X^*, j\in Y^*\}}(f_{ji} - f_{ij})=f_{max}>0$. But 
$r_{ij}=0$ implies $f_{ji}-f_{ij}=-c_{ij}$, so 
$\sum_{\{(ij)\in A|i\in X^*, j\in Y^*\}}(f_{ji} - f_{ij})
=-\sum_{\{(ij)|i\in X^*, j\in Y^*\}}c_{ij}\leq 0$, and therefore a 
contradiction. (In fact the inequality is strict, 
$\sum_{\{(ij)|i\in X^*, j\in Y^*\}}c_{ij}>0$, because $f_{max}>0$, 
and $c_{ij}$ and $c_{ji}$ are simultaneously zero 
or strictly positive.)\\
$b)(s\in X^*$, $t\in X^*$). In this case the flow conservation implies that 
the net flow into $Y^*$ must be zero, i.e. 
$\sum_{\{(ij)\in A|i\in X^*, j\in Y^*\}}(f_{ij} - f_{ji})=0$. Therefore 
$\sum_{\{(ij)\in A|i\in X^*, j\in Y^*\}}r_{ij}
=\sum_{\{(ij)\in A|i\in X^*, j\in Y^*\}}c_{ij}=0$ and 
$\sum_{\{(ij)\in A|i\in X^*, j\in Y^*\}}r_{ji}
=\sum_{\{(ij)\in A|i\in X^*, j\in Y^*\}}c_{ji}>0$. As a result $\exists$ 
$(ij)\in A$ with $c_{ij}=0$ and $c_{ji}>0$, which is a contradiction. 
The case $c)$ is similar to $b)$. 
From all the above it follows that $s\in X^*$ and $t\in Y^*$ so 
$(X^*,Y^*)$ is a cut. 

Now we have to prove that $(X^*,Y^*)$ is also a minimum cut. We have 
$\sum_{\{(ij)\in A|i\in X, j\in Y\}}r_{ij}=0$ and therefore
\begin{equation}
c(X,Y)=\sum_{\{(ij)\in A|i\in X, j\in Y\}}c_{ij}=
\sum_{\{(ij)\in A|i\in X, j\in Y\}}(f_{ij} - f_{ji})=f_{max}
\end{equation}
by the conservation of the flow. This implies, according to the 
Ford-Fulkerson theorem, that $(X,Y)$ is a minimum cut.

\underline{{\it Proposition 3}}: At the end of the algorithm 
${\cal A}_{sg}={\cal A}_{mc}$.

{\it Proof}: ${\cal A}_{mc}\subset{\cal A}_{sg}$ follows from the 
construction of the algorithm and Propositions 1 and 2. In order to prove 
${\cal A}_{sg}\subset{\cal A}_{mc}$ let us assume, without any loss 
of generality, that the supergraph has a single cluster (multiple clusters 
are independent of each other). Let $(IJ)$ be an arc of ${\cal G(N, A)}$ 
connecting subclusters $I$ and $J$. ( We only consider the non-trivial 
case $I\neq Y_t$ and $J\neq X_s$; if $I=Y_t$ or $J= X_s$ the result 
follows immediately from the construction 
of the $S$ cut and $T$ cut.) Let us define:
\begin{equation}
{\cal Y}_{J}=\{\mbox{{\it all subclusters that can be reached from $J$ 
along a directed path in ${\cal G(N, A)}$}}\}
\end{equation}
\begin{equation}
{\cal X}_{I}={\cal N}-{\cal }Y_{J}
\end{equation}
Now $I\notin {\cal Y}_{J}$, otherwise ${\cal G(N, A)}$ would contain a 
directed cycle. 
Therefore $I\in {\cal X}_{I}$, so $({\cal X}_{I},{\cal Y}_{J})$ 
is a directed partition in ${\cal G(N, A)}$
and $(IJ)$ connects this directed partition. This partition  
determines a directed partition in $G^{max}_f$ and therefore a minimum cut 
in $G(N,A)$ according to Proposition 2. Then $(IJ)\subset{\cal A}_{mc}$ 
and because $(IJ)$ is arbitrary ${\cal A}_{sg}\subset{\cal A}_{mc}$ follows.

\underline{{\it Proposition 4}}: ${\cal A}_{mc}$ is the set of arcs 
of $G(N,A)$ that will be saturated by all the maximal flows (the 'weak' 
links of the network).

{\it Proof}: Let $(ij)\in{\cal A}_{mc}$ and $G^{max}_f$ the residual graph 
for a maximal flow. By the definition of ${\cal A}_{mc}$ there exists 
a minimum cut $(X,Y)$ such that $i\in X$ and $j\in Y$. This minimum cut is 
a directed cut in $G^{max}_f$ according to Proposition 2, so $r_{ij}=0$, 
i.e. the arc $(ij)$ is saturated. Conversely, let the arc $(ij)$ 
with capacity $c_{ij}$ be saturated by all the possible maximal 
flows through the 
network, that is flows with value $f=f_{max}$. Let us assume now that 
$(ij)$ is not part of any minimum cut. Therefore, if we decrease the capacity 
$c_{ij}$ by a small value $\epsilon>0$, 
$c_{ij}\rightarrow c^*_{ij}=c_{ij}-\epsilon>0$, 
the maximum flow through the network will still have value $f_{max}$, 
according 
to the Ford-Fulkerson theorem. However, none of the maximal flows will now 
'fit' through the network, in particular through the arc $(ij)$. This is a 
contradiction, therefore the arc $(ij)$ must be part of a minimum cut, 
that is $(ij)\in{\cal A}_{mc}$.

\begin{figure}
\caption{Simplified representation of the network after the first step of 
the algorithm, showing the $S$ cut, the $T$ cut and $Z$ 
(we denote $X_s$ by $S$ and $Y_t$ by $T$ - see text). Note that a 
directed arc connecting $S$ and $T$ may or may not exist.}
\label{1step}
\end{figure}

\begin{figure}
\caption{Simplified representation of the network after the clusters 
making up $Z$ have been identified and an example of a minimum cut.}
\label{2step}
\end{figure}

\begin{figure}
\caption{The supergraph ${\cal G(N, A)}$ at the end of the algorithm and 
an example of a minimum cut.}
\label{3step}
\end{figure}

\begin{figure}
\caption{Ground state structure of the bimodal RFIM for $h/J=3/2$. The spins 
frozen 'up' are green, the spins frozen 'down' are white, while the other 
colors represent the spins making up $Z$; neighbouring subclusters have 
different colors (see text). Note that $Z$ is made up of independent clusters 
that contain one or more subclusters. A dot indicates a field 'up' and 
the absence of it indicates a field 'down'.}
\label{rfim}
\end{figure}

\begin{figure}
\caption{Ground state structure of the DAFF for $h/J=7/2$ and $p=0.9$. 
The color coding is the same as for the RFIM, only now black indicates 
an empty ('nonmagnetic') site. Note that in this case 'flipping' a 
(sub)cluster means coloring it with one or the other of the checkered 
patterns.}
\label{daff}
\end{figure}

\begin{figure}
\caption{A RFIM cluster from Fig. \ref{rfim} and its subcluster graph 
representation; $R$ stand for red, $Y$ for yellow and $B$ for blue. 
The associated degeneracy is $7$.}
\label{cluster}
\end{figure}

\begin{figure}
\caption{Interface configuration for the BDIM at $p=0.64$. The spins 
frozen 'up' are green, the spins frozen 'down' are gray and the 
pieces that are not part of the percolating cluster (see text) are white.  
The other colors represent again the spins making up $Z$, with neighboring 
subclusters having different colors. Note that for this particular sample 
$Z$ is made up of six independent clusters.}
\label{int1}
\end{figure}

\begin{figure}
\caption{A different sample than in Fig. 6, that has three independent 
interfaces, each with its own excitations. Note that $Z$ is in this case 
a single cluster in the algorithm denomination.}
\label{int2}
\end{figure}


\begin{references}
\bibitem{amend}This is true if the average connectivity of a node is 
$O(1)$ as $n(N)\rightarrow\infty$, where $n(N)$ is the number of nodes in $N$.
\end{references}
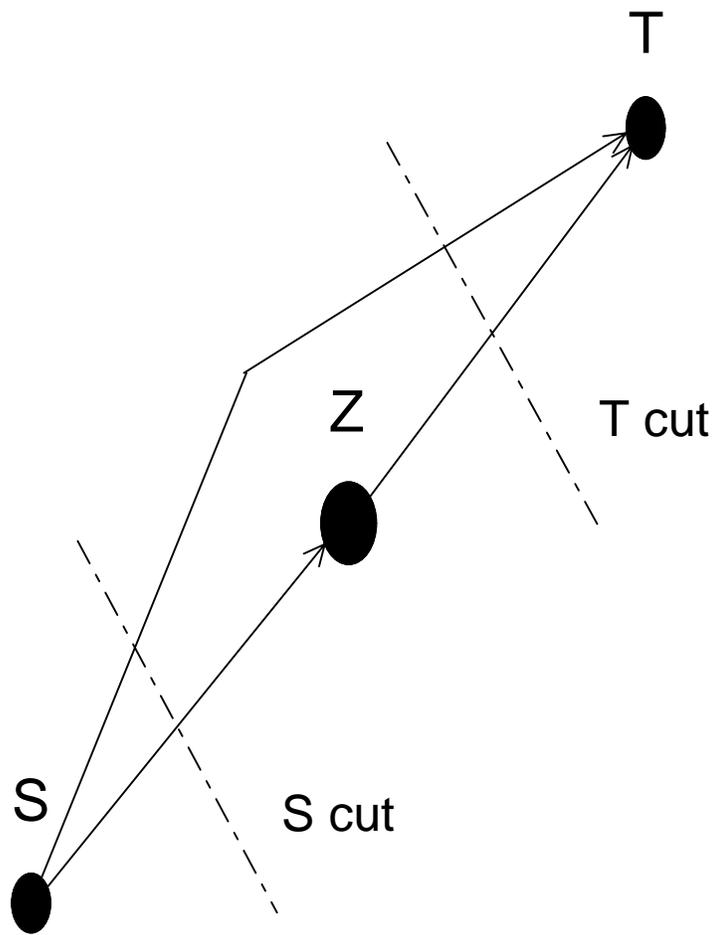
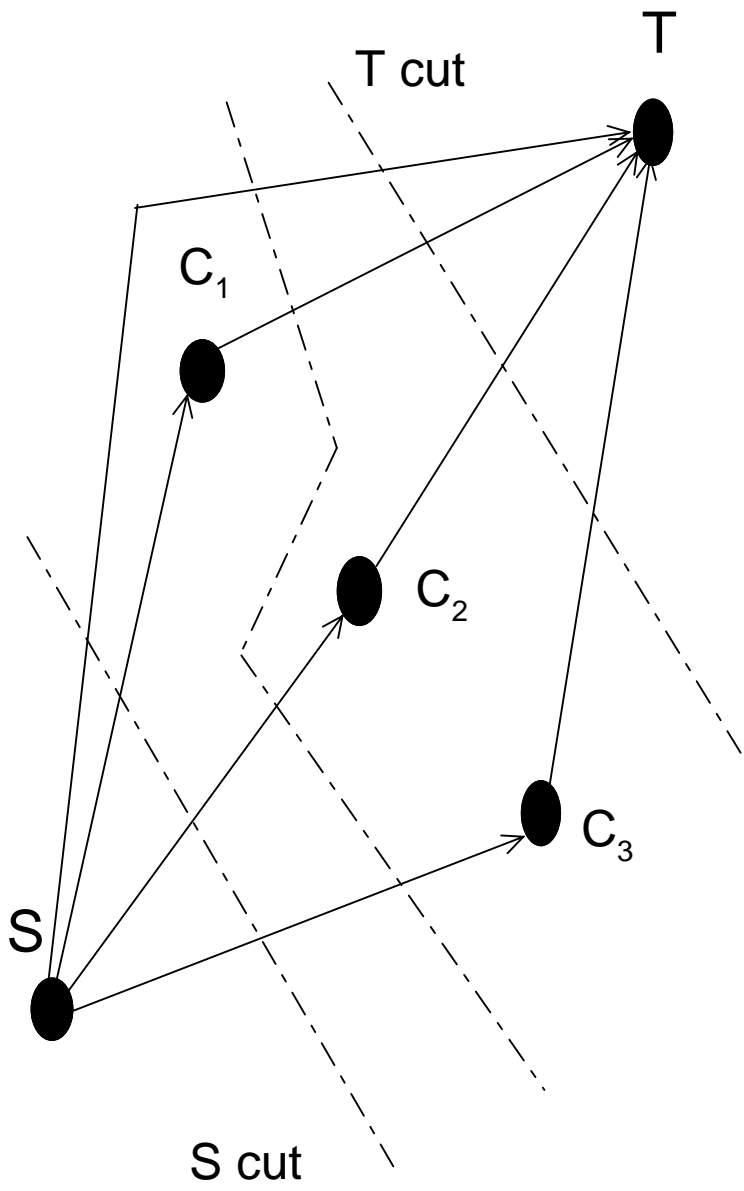
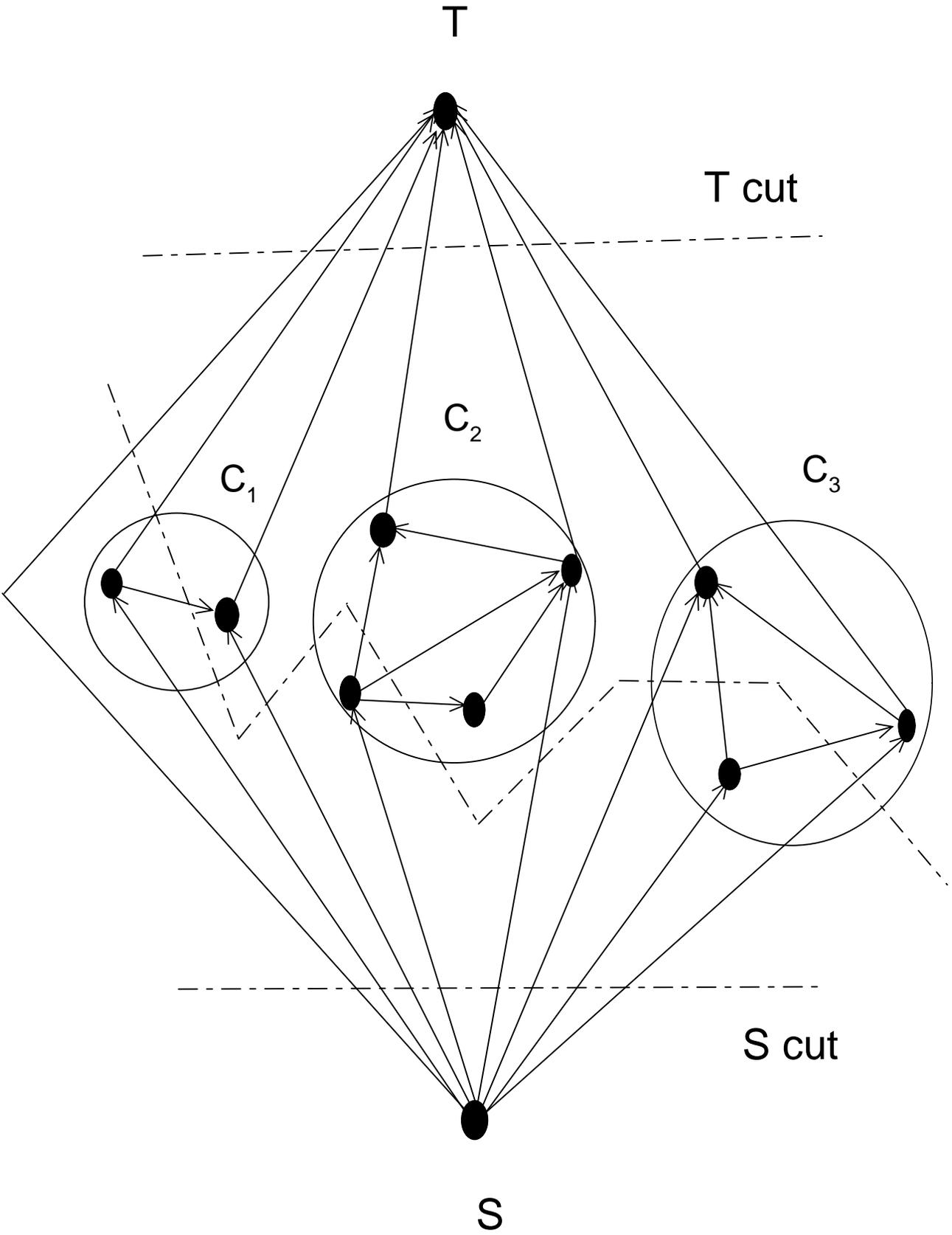
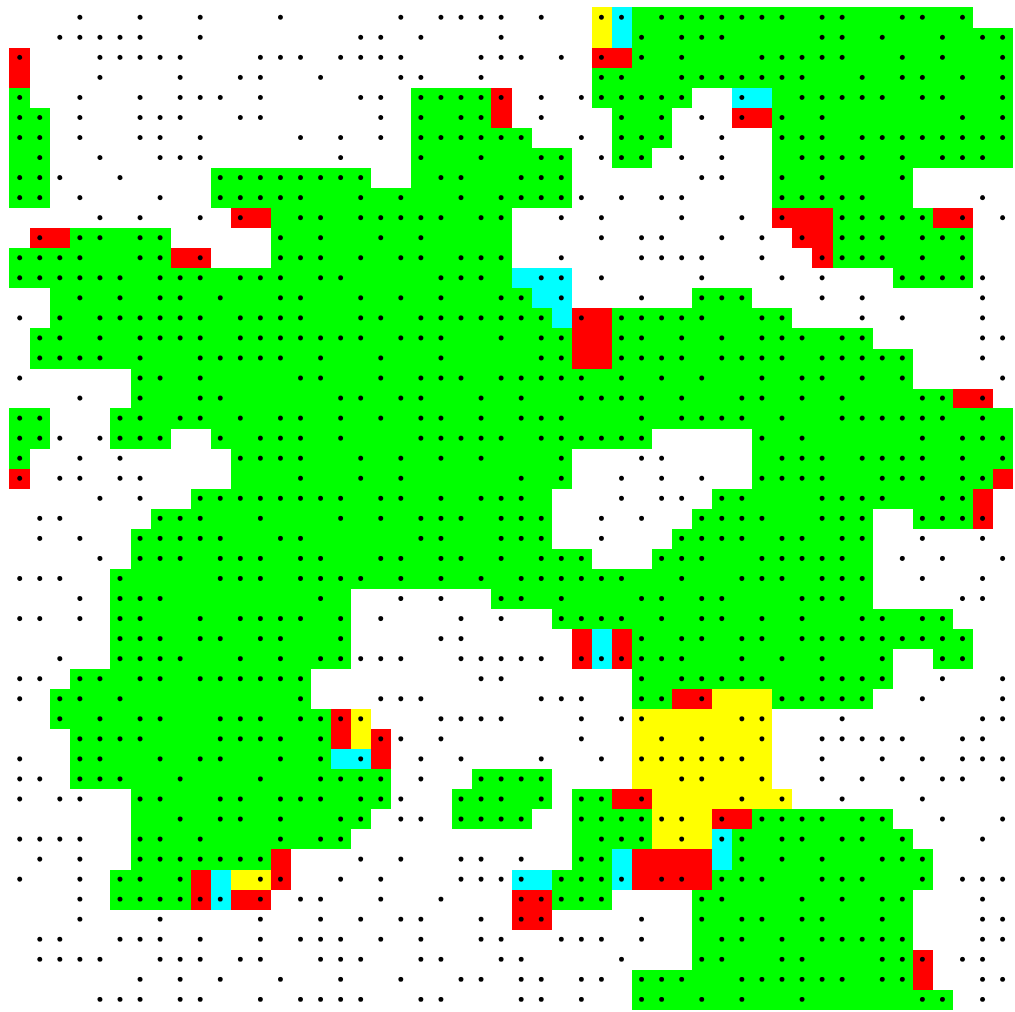
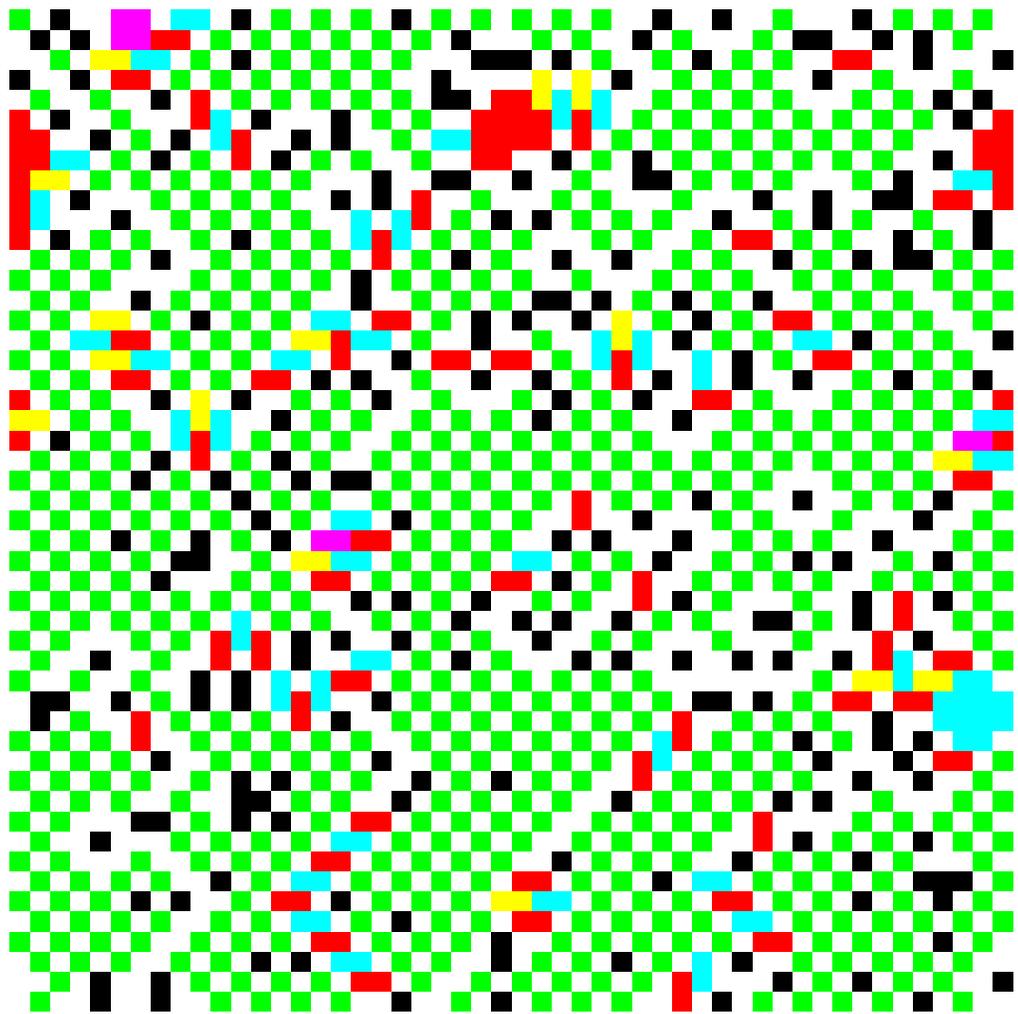
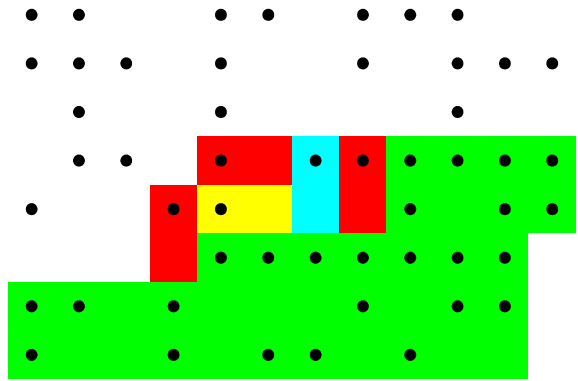
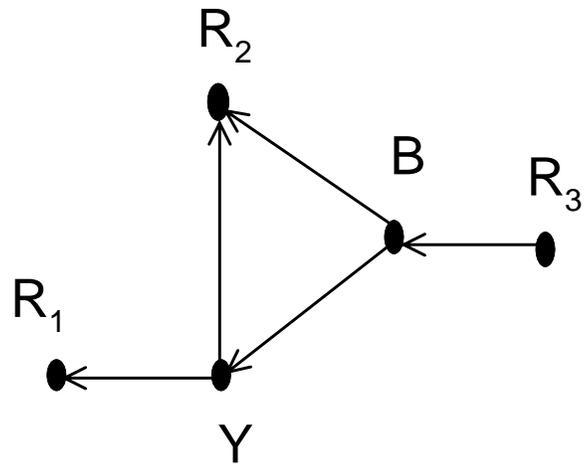
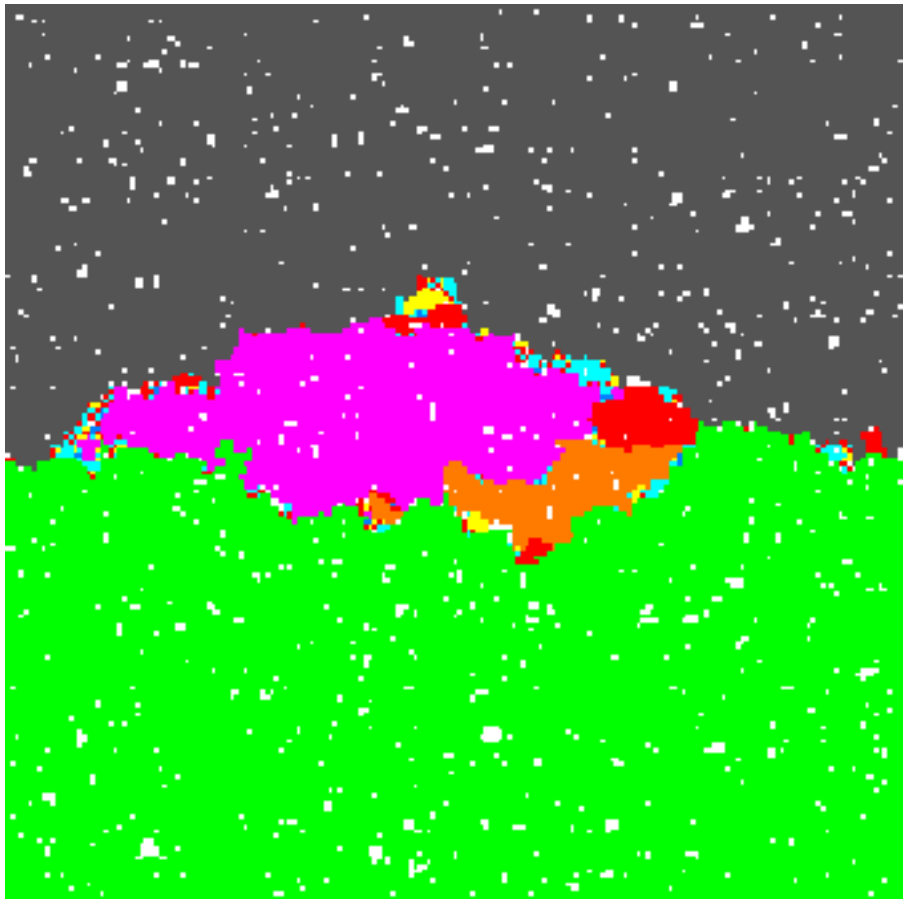
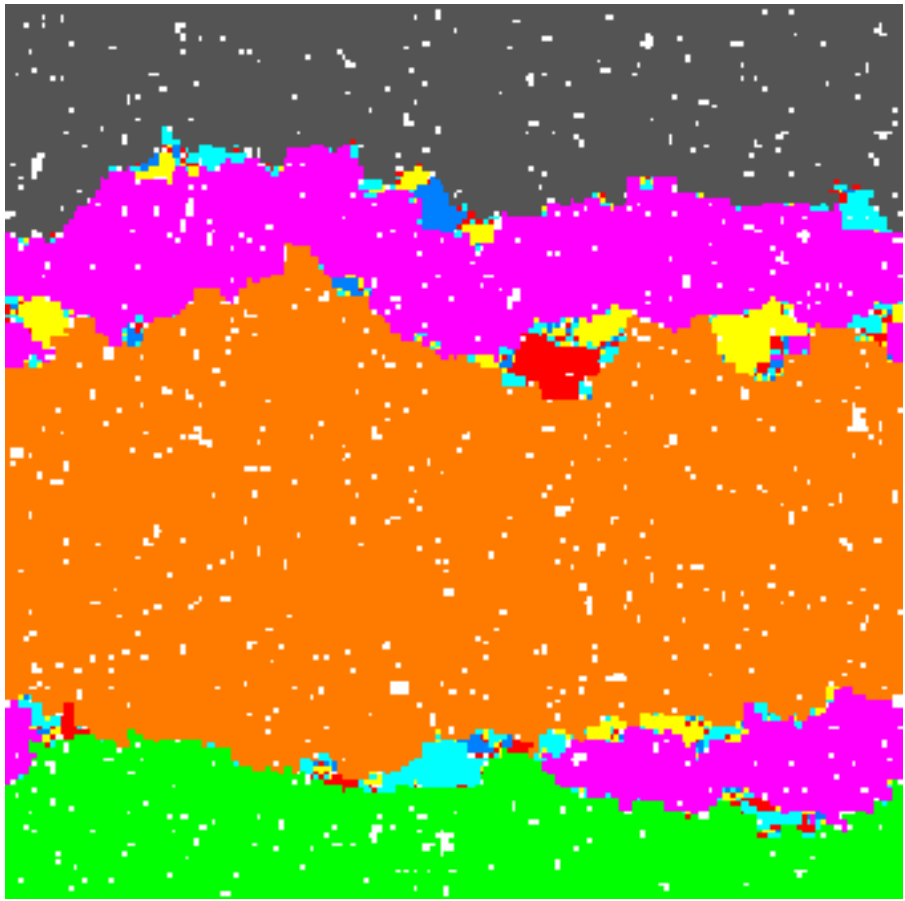
\end{document}